\documentclass[aps,prd,a4paper,superscriptaddress,nofootinbib,
showpacs,showkeys,amsfonts,amssymb,amsmath,11pt]{revtex4}
\usepackage{amssymb,latexsym}
\usepackage{amsmath, amsthm}
\usepackage{amscd}
\usepackage{amsthm}
\usepackage{times}
\usepackage{epsfig}
\usepackage{psfrag}
\usepackage{graphicx}
\usepackage{amssymb,latexsym}

\begin{document}
\title{A Note on Genericity and Stability of Black Holes and 
Naked Singularities in Dust Collapse}
\author{Ravindra V. Saraykar} \email{ravindra.saraykar@gmail.com}
\affiliation{Department of Mathematics, R.T.M. Nagpur 
University, Nagpur, 440033, India}
\author{Pankaj S. Joshi} \email{psj@tifr.res.in}
\affiliation{Tata Institute of Fundamental Research, 
Homi Bhabha road, Colaba, Mumbai 400005, India}

\begin{abstract} 
We comment here on the results in Ref
[4] that showed naked singularities in dynamical gravitational
collapse of inhomogeneous dust to be stable but non-generic. The
definition of genericity used there is reconsidered. We point out
that genericity in terms of an open set, with a positive
measure defined suitably on the space of initial data, is
physically more appropriate compared to the dynamical systems
theory definition used in [4] which makes both black holes and
naked singularities non-generic as collapse outcomes.
\end{abstract}


\maketitle
 
In an early discussion, Hawking [1] emphasized the
importance of the stability and genericity aspects for spacetime
properties in gravitation theory. He pointed out that stable
properties have a special significance in a physical theory when
considering the correspondence between certain physical
observations and a mathematical model. For general relativity the
model is a spacetime, a four-dimensional manifold with a certain
Lorentz metric, satisfying Einstein equations. The accuracy of
real world observations is always limited by practical
difficulties and by the uncertainty principle. Thus only
properties of spacetime which are physically significant are those
that are stable in an appropriate topology and unstable properties
would not have physical relevance.

As the theory of gravitational collapse in general relativity
evolved over a period of past four decades, it is known now that
dynamical gravitational collapse of a massive matter cloud can end
in either a black hole or a naked singularity final state, for
spherical spacetimes with a variety of matter fields and also in
many non-spherical models. Gravitational collapse has been studied
by many authors in detail (see for example [2] and references
therein). Existence of black hole (BH) or naked singularity (NS)
as endstates of collapse is obtained depending on the choice of
initial data from which the collapse evolves, as was shown by
Joshi and Dwivedi (see for example [3]).

A natural question which then arises is, whether these occurrences
and the collapse outcomes in terms of BH or NS final states are
stable and generic with respect to the regular initial data on an
initial spacelike surface from which the gravitational collapse
develops.

From such a perspective, in the case of inhomogeneous dust,
Saraykar and Ghate [4] showed that the occurrence of NS and BH is
stable with respect to small variations in initial data functions
in the sense that initial data set leading the collapse to NS (or
BH) forms an open subset of the full initial data set under a
suitably defined topology on the space of initial data. The
authors assumed there the definition of genericity as given in the
theory of dynamical systems (see e.g. [5]), and it was then argued
that the NS occurrence is stable but not generic as per that 
definition. It is to be noted, however, that the occurrence 
of black holes also then turns out to be non-generic according 
to such a criterion. Therefore, the definition of genericity, 
as used in the dynamical systems studies would not be adequate 
to be used for discussing
the gravitational collapse outcomes. An important point here is,
work of different researchers over a period of last two decades
has shown that the class of initial data set which leads to NS is
disjoint and `fully separated' from that which leads to BH. This
is true for radiation collapse, inhomogeneous dust collapse and
also for general type I matter fields ([2,4,6,7]).

Therefore neither black holes nor naked singularities can 
be dense in the full space of the initial data, and it is clear 
that a different criterion is necessary to access the genericity 
of these collapse final states.
To examine this point, we consider here the situation of the
inhomogeneous dust collapse following [3,4]: The spacetime metric
in the case of inhomogeneous dust is given by
\begin{equation}
    ds^2=-dt^2+\frac{{R'}^2}{1+f(r)}dr^2+R^2d\Omega^2 \; ,
\end{equation}
and the energy momentum tensor and field equations are,
\begin{equation}
    T^{ij} = \rho{\delta^i}_t{\delta^j}_t \; ,
\end{equation}
\begin{equation}
   \rho(t,r) = \frac{F'}{R^2R'}
\end{equation}
\begin{equation}
    \dot{R}^2 = {\frac{F(r)}{R}} + f(r)
\end{equation}
where $T^{ij}$ is the energy-momentum tensor, $\rho$ is the total
energy density and $F(r)$ and $f(r)$ are arbitrary functions of
$r$. The dot denotes a derivative with respect to time, while a
prime denotes a derivative with respect to $r$. Integration of
equation $(4)$ gives
\begin{equation}
 t - t_{0}(R) = \frac{-R^{\frac{3}{2}}G
(\frac{-fR}{F})}{\sqrt{F}}
\end{equation}
where the function $G(x)$ takes value  2/3 at $x = 0$, and
is expressed as inverse sine and inverse hyperbolic sine for other
ranges of $x$ (see e.g. [3,4] for exact expressions for $G(x)$.
Following the root equation method of [3], the condition
for existence of a naked singularity or black hole is given in
terms of the function $\Theta_u(r)$ described as follows ([4]):
\begin{equation}
\Theta_{u}(r) =
\frac{1}{\sqrt{g}(3r^{2}g+r^{3}g^{'})}[\frac{(\frac{Q^{'}}{Q}-
\frac{g^{'}}{g})}{\sqrt{1+\frac{Q}{G}}}
+ (\frac{g^{'}}{g}-\frac{3Q^{'}}{2Q})G(\frac{-Q}{g})]
\end{equation}
where the mass function $F(r)$ and energy function $f(r)$ are
written as
\begin{equation}
F(r) = r^3g(r)
\end{equation}
and
\begin{equation}
f(r) = r^2Q(r)
\end{equation}
Moreover $g(r)$ and $Q(r)$ are sufficiently smooth 
(at least $C^1$)
functions satisfying regularity and energy conditions. 
It is assumed that $g(r)$ satisfies
(i) $g(r) > 0$ and (ii) $rg'(r)+ 3g > 0$. These are 
positivity of mass and energy conditions.

Then, if the value of the function $\Theta_{u}(r)$ at $r = 0$ 
is greater than $\alpha$ where,
\begin{equation}
\alpha = \frac{13}{3} + \frac{5}{2} \times \sqrt{3},
\end{equation}
then the tangent to a nonspacelike curve will be positive, {\it
i.e.} future directed nonspacelike curves will reach the
singularity in the past. In other words, singularity will be
naked, not covered by an event horizon. If this condition is
reversed, we get a black hole.

We now consider \emph{A} to be the class of all continuous 
functions $A(r)$ defined on $[0,r_b]$. Consider a subclass of 
\emph{A},
denoted by $\emph{A}_1$, consisting of functions $A(r)$, such
that $ A(0) > \alpha$. Now consider the equation
\begin{equation}
\Theta_u(r) = \emph{A(r)},
\end{equation}
regarded as differential equation in
$Q(r)$) where $\ \Theta_u(r) $ is given by (6).

Then the following result was proved in [4] :
Given a function $g(r)$ satisfying the above conditions, there are infinitely
many choices of function $A(r)$ in the class $\ \emph{A}_1 $ such that
for each such choice of
$A(r)$, there exists a unique function $Q(r)$ such that the initial
data ${(g(r), Q(r))}$
leads the collapse to a naked singularity. Thus, the conditions
on $g(r)$ and $A(r)$ leading the
collapse to NS are, 

(i) $g(r) > 0$, (ii) $rg'(r)+ 3g > 0$ and (iii) $\ A(0) > \alpha$.

Since the existence of $Q(r)$ is guaranteed by $g(r)$ and $A(r)$, we
can consider the set $\emph{N}$
of all ${(g(r), A(r))}$ (instead of ${(g(r), Q(r))}$), satisfying the
above conditions, as the set of initial data leading the collapse to
NS. As proved in [4], this set forms an open subset of
$\ \emph{G}\times\emph{A} $ where $\ \emph{G}$ is the space
of all $C^1$ functions defined on the interval $[0,r_b]$, 
and \emph{A} is as above. We note that since the consideration 
of function $A(r)$ comes from equation (6) which contains 
$g(r)$, $Q(r)$ and their first derivatives, it is sufficient to 
assume that the functions $A(r)$ are continuous.

Similarly, the set $\ \emph{B}$ of all ${(g(r), A(r))}$ 
satisfying the conditions,

(i) $g(r) > 0$, (ii) $rg'(r)+ 3g > 0$, and (iii) $\ A(0) < \alpha$,

will lead the collapse to a black hole, and similar arguments
imply that $\ \emph{B}$ is also
an open subset of $\ \emph{G}\times\emph{A}$. In this sense, 
both these BH and NS occurrences in collapse are stable.
It is also clear that $\ \emph{N}$ and $\emph{B}$ are disjoint.
Therefore, none of them can be dense in $\ \emph{G}\times\emph{A} $.
Nevertheless, each of these sets are substantially big and 
it can be shown that they have a non-zero positive 
measure [8].

It is thus clear that if we follow strictly the dynamical 
systems definition of genericity, then both the outcomes of 
collapse, namely NS and BH would be non-generic. Thus, it is
reasonable to argue that a change in the definition of genericity
is desirable. This change is also justified by the work of other
relativists who used the nomenclature 'generic' in the sense of
'abundance' or existence of an open set of non-zero measure
consisting of initial data leading the collapse to BH or NS as in
the case of scalar fields or for AdS models (see {\it e.g.} 
[9, 10] and references therein, but see also [11] where genericity
is defined in terms of codimension). We note that in general 
it is clear from the definitions of `stability' and `genericity' used
that these concepts depend upon the topologies under consideration. 
Thus a given property may be stable and generic in
some topologies and not so in others. Which of the topologies is
of physical interest will depend upon the nature of the property
under consideration.

It follows that the dynamical systems definition of `genericity'
needs to be modified if we desire to have black holes as generic
outcomes of gravitational collapse of dust. Dust collapse is
clearly one of the most fundamental collapse scenarios, as the
classic Oppenheimer-Snyder homogeneous dust collapse model is at
the very foundation of the modern black hole physics and its
astrophysical applications.

We could therefore formulate an appropriate and physically 
reasonable criterion of genericity for the dust collapse outcomes 
as follows:

We assume that the collapse begins with a regular initial 
data, with weak energy condition and other regularity conditions 
satisfied, {\it e.g.} that there are no shell crossings $R'=0$ 
as the collapse evolves. The initial data, namely $F(r)$ and 
$f(r)$ (or $g(r)$ and $A(r)$) allow for the formation of 
both black holes and naked singularities, and we call each 
of these outcomes to be generic if the following 
conditions are satisfied:

    (i) The set of initial data which evolves the collapse to naked 
singularity (or black hole) is an open subset of the full space of 
initial data.

    (ii) If we impose a positive measure on the space of initial data, 
then the set of initial data with each of these outcomes should have 
a non-zero measure in the total space.

The first condition above means given an initial data point 
$F_1(r)$ and $f_1(r)$, which evolves the collapse to naked singularity 
(black hole), there should be an open neighborhood of this data 
point such that each initial data in this neighborhood also 
evolves to the same outcome. The second condition here means that 
each of these outcomes are substantially big in the full 
space of initial data.

We see from the consideration above that this holds true for 
dust collapse. In other words, an outcome of collapse, either in 
terms of a black hole or naked singularity is called `generic' 
if there exists a subset or region of the initial data space 
that leads the collapse to such an outcome, and which has 
a positive measure. Then actually how `big' such a region or
the subspace of the initial data would be, depends on the 
collapse model being considered. For example, for the Vaidya 
radiation collapse, each of the regions going to BH or NS 
seem to be both finitely big and with a non-zero measure, and 
for dust collapse also they seem to have essentially 
equal 'sizes'.

Thus, we observe the following: With the above definition 
of genericity, we find that both the outcomes of dust collapse, 
namely black holes and naked singularities are generic 
and also stable (see also [8] for a recent discussion on 
perfect fluid collapse). Since denseness depends upon the 
parent set chosen as well as the choice of topology, choosing 
the definition such as above looks physically reasonable. 
In other words, it is reasonable to argue that a change in 
the definition of genericity which will make both these
collapse final states generic is desirable.

As we are aware, the concepts such as `stability' and `genericity'
which are so important for physical considerations, are not
well-defined in the Einstein gravitation theory, unlike the
Newtonian case. This is of course the crux of the problems associated
with any precise formulation of the cosmic censorship conjecture.
It is therefore clear that a deeper, more detailed and if possible
case by case consideration of collapse models may help understand
these aspects better. In this spirit, we believe the consideration
above provides useful insights on cosmic censorship, which
continues to be one of the most outstanding and important problems
in gravitation theory today. The stability and genericity for
collapse outcomes in the context of a general gravitational
collapse will be discussed elsewhere.
\vspace{10mm}

{\bf References}

1. S.W.Hawking, Gen. Rel. Grav. 1, 393 (1970).

2. P.S.Joshi, Global aspects in gravitation and cosmology, Oxford Press, 1993.
P.S.Joshi, Gravitational collapse and spacetime singularities,
Cambridge press, 2007.

3. I.H.Dwivedi and P.S.Joshi, Class.Quant.Grav. 14 (1997) 1223.

4. R. V. Saraykar and S. H. Ghate, Class. Quantum Grav. 16, 281 (1999).

5. R. Abraham and J.E.Marsden, Foundations of Mechanics, 2nd edition,
Addison-Wesley (1978)

6. S. B. Sarwe and R. V. Saraykar, Pramana 65, 17 (2005).

7. I.H.Dwivedi and P.S.Joshi,Phys. Rev.D 47, 5357 (1993)

8. P.S.Joshi, D.Malafarina and R.V.Saraykar,  arXiv.org:1107.3749
[gr-qc]; to be published in Int.J.of Mod.Phys. D (2012).

9. C. Gundlach, Phys. Rept. 376, 339 (2003); C. Gundlach and
J. M. Martin-Garcia, \\ arXiv:0711.4620v1 [gr-qc];

10. T. Hertog, G.T. Horowitz, K. Maeda, Phys.Rev.Lett.92,131101 (2004),
also arXiv.org:gr-qc/0405050 v2

11. D. Christodoulou, Ann. of Math. \textbf{149}, 183 (1999).

\end{document}